\documentstyle[12pt,twoside]{article}
\input amssym.def   
\input amssym.tex

\def\beq{\begin{equation}}
\def\eeq{\end{equation}}
\def\bi{\begin{itemize}}
\def\ei{\end{itemize}}
\def\beqar{\begin{eqnarray}}
\def\eeqar{\end{eqnarray}}
\newcommand{\rmd}{{\rm d\null}}
\newcommand{\dotrho}{\dot{\rho}}
\newcommand{\dotv}{\dot{v}}

\newcommand{\dotth}{\dot{\theta}}
\newcommand{\dotpsi}{\dot{\psi}}
\newcommand{\sql}{\sqrt{2\lambda}}

\newcommand{\ppp}{\psi\partial_x\psi}

\makeatletter
\@addtoreset{equation}{section}
\makeatother

\begin{document}

\title{Integrable Supersymmetric Fluid Mechanics from Superstrings}
\author{Y. Bergner and R. Jackiw \\
\small\it Center for Theoretical Physics\\ 
\small\it Massachusetts Institute of Technology\\
\small\it Cambridge, MA ~02139--4307, USA}
\date{\small MIT-CTP-3106 \quad  Submitted to Phys. Lett. A}
\maketitle

\begin{abstract}%
Following the construction of a model for the planar supersymmetric Chaplygin gas, supersymmetric fluid mechanics in (1+1)-dimensions is obtained from
the light-cone parametrized Nambu-Goto superstring in (2+1)-dimensions. The lineal model is completely
integrable and can be formulated neatly using Riemann coordinates.  Infinite towers of conserved charges and supercharges are exhibited.  They form irreducible representations of a dynamical (hidden) $SO(2,1)$ symmetry group.
\end{abstract}

\section{Introduction} 

The Galileo invariant equations governing isentropic fluids in one
spatial dimension (continuity and Euler equations for the density
$\rho$ and velocity $v$) are completely integrable for polytropic
gases (pressure $\propto \rho^n$) \cite{multi}, and are accompanied by
the usual hallmarks of complete integrability: Lax pairs, infinite
number of constants of motion, etc. \cite{Nutku}  Especially interesting is the
Chaplygin gas ($n=-1$) because this model possesses the further hidden
symmetry of (2+1)-dimensional Poincar\'e invariance, which is a
consequence of the fact that this fluid model in (1+1)-dimensional
spacetime devolves from the Nambu-Goto model for a string moving on
the plane, after the parameterization invariance of the latter is
fixed \cite{comment}.

In this Letter we enlarge the lineal Chaplygin gas to include
anti-commuting Grassmann variables, so that the extended model
supports a supersymmetry.  This is achieved by considering a
superstring moving on a plane and again fixing the parameterization
invariance. The construction is analogous to what has already been
done in one higher dimension:  the Nambu-Goto action for a
supermembrane in (3+1)-dimensions gives rise, in a specific
parameterization, to a supersymmetric planar Chaplygin gas \cite{RJAPPRD}.  
%These two models, the lineal and planar supersymmetric Chaplygin fluid, 
Lineal and planar supersymmetric fluid models appear to be the
only possible examples of the supersymmetric Nambu-Goto/fluid connection.  For a higher
dimensional generalization, the reduction program would begin with a
$p$-brane in $D=p+2$ spacetime, giving rise to a fluid in
$D=p+1$ spacetime.  While there are no constraints on $p$ in the
purely bosonic case, supersymmetric extensions are greatly
constrained: the {\em brane-scan} for ``fundamental'' 
super $p$-branes (i.e. with only scalar supermultiplets in the
worldvolume) contains only the above two cases cases \cite{Duff},
$p=2$ in $D=4$ and $p=1$ in $D=3$.
As we demonstrate, the supersymmetric extension enjoys the same
integrability properties as the purely bosonic, lineal Chaplygin gas,
as a consequence of the complete integrability for the dynamics of the
superstring on the plane.

\section{Superstring Formulation}

We begin with the Nambu-Goto superstring in $D=3$ 
%\cite{Gauntlett}
:

\beq
\label{stringaction}  I = -\int \rmd \tau \rmd \sigma \, \{\sqrt{g} - i
\epsilon^{ij} \partial_i X^{\mu} \bar{\psi} \gamma_{\mu} \partial_j
\psi \},
\eeq
where
\beqar
g &=& -\mbox{det} \{\Pi^{\mu}_{i}\Pi^{\nu}_{j}\eta_{\mu\nu} \},\\
\Pi^{\mu}_{i} &=& \partial_{i}X^{\mu} - i \bar{\psi} \gamma^{\mu}
\partial_i \psi \ .
\eeqar
In these expressions $\mu,\nu$ are spacetime indices running over $0,
1, 2$ and $i,j$ are worldsheet indices denoting $\tau$ and $\sigma$.  We now go to the light-cone
gauge where we define $X^{\pm}=\frac{1}{\sqrt{2}}(X^0 \pm X^2)$.
$X^{+}$ is identified with the timelike parameter $\tau$, $X^{-}$ is
renamed $\theta$, and the remaining transverse component $X^1$ is renamed $x$.  We
can choose a two-dimensional Majorana representation for the
$\gamma$-matrices: 
$$\gamma^0=\sigma^2, \ \ \gamma^1= -i\sigma^3, \ \ \gamma^2= i \sigma^1,$$
 such that $\psi$ is a real, two-component spinor.  A remaining
fermionic gauge choice sets  
$$\gamma^{+} \psi=0,$$
where $\gamma^{\pm}=\frac{1}{\sqrt{2}}(\gamma^0 \pm \gamma^2)$.  Thus $\psi$
is further reduced to a real, one-component Grassmann field.  Finally we define
the complex conjugation of a product of Grassmann fields $(\psi_1
\psi_2)^\star = \psi^{\star}_{1}\psi^{\star}_2$ so as to eliminate $i$
from Grassmann bilinears in our final expression.  The light-cone
gauge-fixed Lagrange density becomes:
\beq
{\cal L} = -\sqrt{g\Gamma} + \sqrt{2} \psi \partial_{\sigma}\psi,
\eeq
where
\beqar
g&=&(\partial_{\sigma}x)^2, \\
\Gamma &=& 2 \partial_{\tau}\theta - (\partial_{\tau}x)^2 - 2\sqrt{2} \psi \partial_{\tau}\psi +
\frac{u^2}{g},\\
u&=& \partial_{\sigma}\theta - \partial_{\tau}x \partial_{\sigma}x - \sqrt{2}\psi\partial_{\tau}\psi \ .
\eeqar
In the above equations, $\partial_{\sigma}$ and $\partial_{\tau}$
denote partial derivatives with respect to the spacelike and timelike
worldsheet coordinates.  The canonical momenta
\beqar
\label{pdef} p&=&\frac{\partial {\cal L}}{\partial (\partial_{\tau}x)} =
\sqrt{\frac{g}{\Gamma}}( \partial_{\tau}x + \frac{u}{g} \partial_{\sigma} x), \\ 
\Pi&=&\frac{\partial {\cal L}}{\partial (\partial_{\tau}\theta)}= -\sqrt{\frac{g}{\Gamma}},
\eeqar
satisfy the constraint equation
\beq \label{constraint}
p \partial_{\sigma} x + \Pi \partial_{\sigma} \theta - \sqrt{2} \Pi
\psi \partial_{\sigma} \psi = 0
\eeq
and can be used to recast ${\cal L}$ into the form
\beq
{\cal L} = p\partial_{\tau}x + \Pi \partial_{\tau}\theta +
\frac{1}{2\Pi}(p^2+g)+\sqrt{2}\psi\partial_{\sigma}\psi-\sqrt{2}\Pi\psi\partial_{\tau}\psi
+ u(p \partial_{\sigma}x + \Pi \partial_{\sigma}\theta - \sqrt{2} \Pi
\psi \partial_{\sigma} \psi),
\eeq
where $u$ is now a Lagrange multiplier enforcing the constraint.
We use the remaining parameterization freedom to fix $u=0$ and $\Pi=-1$
and perform a hodographic transformation, interchanging independent 
with dependent variables \cite{RJAL}.  The partial derivatives
transform by the chain rule:
\beqar
\partial_{\sigma} &=& (\partial_{\sigma}x) \partial_x = \sqrt{g} \partial_x\ , \\
\partial_{\tau} &=& \partial_t + (\partial_{\tau}x) \partial_x =
\partial_t + v\partial_x \ ,
\eeqar
and the measure transforms with a factor of $1/\sqrt{g}$.  Finally,
after renaming $\sqrt{g}$ as $\sqrt{2\lambda}/\rho$, we obtain the
Lagrangian for the Chaplygin ``super'' gas in (1+1)-dimensions (below and in
what follows the overdot denotes derivative with respect to
time $t$):
\beq
\label{lagr}
L = \frac{1}{\sqrt{2\lambda}} \int {\rm d\null}x \, { \{
 -\rho(\dotth-\frac12\psi\dotpsi)-\frac12\rho v^2 -
\frac{\lambda}{\rho} + \frac{\sqrt{2\lambda}}{2}\psi\partial_x\psi \} }\ ,  
\eeq
where according to (\ref{pdef}) and (\ref{constraint}) (at $u=0$ and $\Pi=-1$)
\beq
\label{vdef} v=p=\partial_x\theta-\frac12 \psi\partial_x\psi \ .
\eeq
We have used $\rho$ and $v$ in anticipation of their role as the
fluid density and velocity, and we demonstrate below that they indeed satisfy
appropriate equations of motion.  For convenience we have also
rescaled $\psi$ everywhere by a factor of $2^{-3/4}$. The Lagrangian
(\ref{lagr}) agrees with the limiting case of the planar fluid in
\cite{RJAPPRD}. 
We note that as for the planar case, a more straightforward derivation
leads to the fluid Lagrangian of (\ref{lagr}) with $\rho$ integrated
out.  Specifically, if the parameterization freedom is used directly to
equate the spacelike and timelike coordinates $\sigma$ and $\tau$ with
$x$ and $t$, we obtain
\beq
L' = - \int {\rm d\null}x \,{\Bigl( \sqrt{2 \dotth - \psi \dotpsi +
v^2} - \frac12 \ppp \Bigr)}\ ,
\eeq
where $v$ is defined as in (\ref{vdef}).  This form of the Lagrangian
can be obtained from (\ref{lagr}) after $\rho$ is eliminated using the
equations of motion for $\theta$ and $\psi$, shown below. 
\section{The Supersymmetric Chaplygin Gas}
\subsection{Equations of Motion}
The following equations of motion are obtained by variation of
the Lagrangian (\ref{lagr}):
\beqar
\label{rhoeq} \dotrho + \partial_x(\rho v) &=& 0, \\
\label{psieq} \dotpsi + \Bigl( v+\frac{\sql}{\rho}\Bigr)\partial_x\psi &=& 0,\\
\label{theq} \dotth +
v\partial_x\theta&=&\frac12v^2+\frac{\lambda}{\rho^2}-\frac{\sql}{2\rho} \ppp, \\
\label{veq} \dotv + v\partial_x v &=& \partial_x\Bigl(\frac{\lambda}{\rho^2}\Bigr) \ .
\eeqar
Naturally, there are only three independent equations of motion as
(\ref{veq}) is obtained from (\ref{psieq}), (\ref{theq}) and
(\ref{vdef}). Equations (\ref{rhoeq}) and (\ref{veq}) are seen to be just the continuity
and Euler equations for the Chaplygin gas.  Note that these do not see
the Grassmann variables directly. 

We now pass to the Riemann coordinates, which for this system are
(velocity $\pm$ sound speed $\sql/\rho$): 
\beq
R_{\pm} = \Bigl(v \pm \frac{\sql}{\rho}\Bigr) \ .
\eeq
In terms of the Riemann coordinates, the equations of motion obtain
the form
\beqar
\label{Req} \dot{R_{\pm}} &=& - R_{\mp}\partial_x R_{\pm}, \\
\label{psiReq}\dotpsi &=& -R_{+}\partial_x\psi, \\
\label{thReq} \dotth &=& -\frac12 R_{+}R_{-}-\frac12 R_{+} \ppp \ .
\eeqar
The equations in (\ref{Req}) contain the continuity and Euler equations and
are known to be integrable \cite{Nutku}.  It is readily verified that
equation (\ref{psiReq}) for $\psi$ is solved by any function of
$R_{-}$, 
\beq
\label{sol}\psi = \Psi(R_{-}),
\eeq
and hence the fluid model is completely integrable.  That this is the case
should come as no surprise considering that we began with an
integrable world-sheet theory.  

At this
point it may seem curiously asymmetric that equation (\ref{psiReq})
for the Grassmann field should contain the $R_{+}$ Riemann coordinate
and not the $R_{-}$ companion coordinate.  In fact, the reverse would have
been the case if the sign of the $\sql$ term in (\ref{lagr}) had
been opposite.  The entire model is consistent with this substitution,
which is just the choice of identifying $\sqrt{g}$ with plus or
minus the sound speed $\sql/\rho$.

The energy-momentum tensor is constructed from (\ref{lagr}), and
its components are
\beqar
\label{hamil} T^{00} &=& {\cal H} = \frac12 \rho v^2 + \frac{\lambda}{\rho} -
\frac{\sql}{2}\ppp, \\
T^{01} &=& {\cal P} = \rho v, \\
T^{10} &=& \frac{\rho v}{2} R_{+}R_{-} - \frac{\sql}{2}R_{+}
\ppp, \\
T^{11} &=& \rho  R_{+}R_{-} \ .
\eeqar
The expected conserved quantities of the system, the generators of the
Galileo group, are verified to be time-independent using the equations
of motion.  We have
\beqar
\label{Neq}N&=&\int \rmd x \, \rho, \\
P&=&\int \rmd x \, \rho v, \\
H&=&\int \rmd x \, \Bigl(\frac12 \rho v^2 + \frac{\lambda}{\rho} -
\frac{\sql}{2}\ppp \Bigr), \\
%&=& H_{0} - \frac{\sql}{2} \int{\rmd x \, \ppp} \equiv H_0 - \frac{\sql}{2}A_0, \\
\label{Beq}B&=&\int \rmd x \, \rho (x-vt) = \int{\rmd x \, x\rho} - t P,
\eeqar
%where it should be noted that $H_{0}$, the Hamiltonian without the
%last term, is 
%still separately time independent. 
Although  some generators look purely bosonic, there are still
Grassmann fields hidden 
in $v$ according to its definition (\ref{vdef}).  

In going to Riemann
coordinates, we can observe a ladder of conserved charges of the form \cite{RJAP1}
\beq
\label{bocharges}I^{\pm}_{n}=\int \rmd x \, \rho R_{\pm}^{n}  \ .
\eeq
The first few values of $n$ above give
\beqar
I^{\pm}_{0} &=& N \\
I^{\pm}_{1} &=& P \pm \sql \Omega \\
I^{+}_{2} &=& 2 H \ , 
%= 2 \Bigl(H_0 \mp \frac{\sql}{2} A_0 \Bigr),
\eeqar
where $\Omega$ is used to denote the length of space $\int \rmd x$.  (Note
that $I^{-}_{2},$ would correspond to the Hamiltonian of the theory
with $\sql$ replaced by its negative).  
%the  plus or minus choice has been absorbed
%into the freedom of defining $H$ to be $H_0 \mp \frac{\sql}{2}A_0$ as
%described above.  

Ref. \cite{RJAPPRD} identified two different supersymmetry generators,
which correspond in one space dimension to the time independent
quantities:
\beqar
\label{Qtildeeq} \tilde Q &=& \int \rmd x \, \rho \psi, \\
\label{Qeq}Q &=& \int \rmd x \, \rho \Bigl(v-\frac{\sql}{\rho}\Bigr) \psi \ .
\eeqar
These are again but special cases ($n=0$ and $n=1$) of a ladder of
conserved supercharges described by
\beq
\label{supercharges}Q_{n} = \int \rmd x \, \rho R_{-}^n \psi \ .
\eeq
We see that the supercharges evaluated on the solution (\ref{sol})
reproduce the form of the bosonic charges (\ref{bocharges}). 
 
Let us observe that there exist further bosonic and fermionic
conserved charges.  For example, one may verify that the bosonic charges
%\beqar\label{Acharges}
%I^{\pm}_{na} = \int \rmd x \, \rho R_{\pm}^n \Bigl(\frac{\partial R_{\pm}}{\rho}\Bigr)^a \ , \\
%K_{na} = \int \rmd x \, \rho R_{-}^n \Bigl(\frac{\partial R_{-}}{\rho}\Bigr)^a \ppp \\
%Q_{na} = \int \rmd x \, \rho R_{-}^n \Bigl(\frac{\partial R_{-}}{\rho}\Bigr)^a \psi \\
%S_{na} = \int \rmd x \, R_{-}^n \Bigl(\frac{\partial R_{-}}{\rho}\Bigr)^a \partial_x \psi \\
%\eeqar
%are all conserved for positive-integer values of the indices $n$ and
%$a$.  
\beqar
\label{Ibeq}\int \rmd x \, \rho R_{\pm}^n \Bigl(\frac{\partial_x
R_{\pm}}{\rho}\Bigr)^m  \\
\label{Aeq} \int \rmd x \, \rho R_{-}^n \Bigl(\frac{\ppp}{\rho}\Bigr)
\eeqar 
are conserved, as are the fermionic charges
\beq\label{Keq} 
\int \rmd x \, \rho R_{-}^n \Bigl(\frac{\partial_x\psi}{\rho}\Bigr).
\eeq
Conserved expressions involving higher derivatives may also be constructed.
The conservation of these quantities is easily understood when
the string worldsheet variables are used.  Then the above are written
as $\int \rmd \sigma R_{\pm}^n (\partial_\sigma R_{\pm})^m$, $\int
\rmd \sigma R_{-}^n (\psi \partial_\sigma \psi)$, and $\int
\rmd \sigma R_{-}^n (\partial_\sigma \psi)$, respectively.  Furthermore when
$R_{\pm}$ are evaluated on solutions, they become functions of $\tau
\pm \sigma$ \cite{Bazeia}, so that integration over $\sigma$ extinguishes the $\tau$
dependence, leaving constant quantities.

%We have
%also obtained another ladder of bosonic charges from the Grassmann
%bilinear: 
%\beq\label{Acharges}
%A_n = \int \rmd x \, R_{-}^n \ppp \ .
%\eeq
\subsection{Canonical Structure}
The equations of motion (\ref{rhoeq}-\ref{theq}) can also be obtained
by Poisson bracketing with the Hamiltonian (\ref{hamil}) if the following
canonical brackets are postulated:
\beqar
\lbrace \theta(x),\rho(y) \rbrace &=& \delta(x-y),\\
\lbrace \theta(x),\psi(y)\rbrace &=&-\frac{\psi}{2\rho}\delta(x-y),\\
\lbrace \psi(x),\psi(y) \rbrace &=&-\frac{1}{\rho}\delta(x-y), 
\eeqar
where the last bracket, containing Grassmann arguments on both sides is
understood to be the anti-bracket. 
With these one verifies that the conserved charges in
(\ref{Neq})-(\ref{Beq}) generate the appropriate Galileo symmetry
transformations on the dynamical variables $\rho$, $\theta$, and
$\psi$.  Correspondingly the supercharges (\ref{Qtildeeq}),(\ref{Qeq})
generate the super transformations

%The algebra of the bosonic generators contains the nonzero brackets
%\beqar
%\{B,H\} &=& -P , \\
%\{B,P\} &=& -N ,
%\eeqar
%where $B$ is the generator of Galileo boosts.  The remaining brackets
%all vanish.
% The supercharges $Q$ and
%$\tilde Q$, when bracketed with the fields, are seen to effect the
%following transformations (in terms of a Grassmann parameter $\eta$):
\beqar
\tilde\delta\rho&=&0 \qquad\qquad\quad \, \delta\rho = -\eta \partial_x(\rho\psi) \\
\tilde\delta\theta &=& -\frac12 \eta \psi \qquad\quad \delta\theta =
-\frac12 \eta R_{+} \psi  \\
\tilde\delta\psi&=&-\eta \qquad\qquad \, \delta\psi = -\eta \ppp - \eta R_{-}
%\delta\rho &=& -\eta (\rho\psi)' \\
%\delta\theta &=& -\frac12 \eta R_{+} \psi \\
%\delta\psi &=& -\eta \psi \psi' - \eta R_+
\eeqar
which leave the Lagrangian (\ref{lagr}) invariant.
The algebra of the bosonic generators reproduces the algebra of the
(extended) Galileo group, the extension residing in the bracket
$\{B,P\} = -N$. The algebra of the supercharges is
\beqar
\{\bar{\eta} Q, \eta Q\} &=& 2 \bar{\eta}\eta   H\\
\{\bar{\eta} \tilde Q, \eta \tilde Q\} &=&  \bar{\eta} \eta N \\
\{\bar{\eta} \tilde Q, \eta Q\} &=& \bar{\eta} \eta(P - \sqrt{2\lambda}  \Omega) \\
\{B, Q\} &=& \tilde Q \ .
\eeqar

%Finally we record the transformation effected on the fields by $A_0$
%from the second tower of bosonic charges.  We find that
%\beqar
%\delta\rho &=& 0 \\
%\delta\theta &=& \frac{1}{\rho} \ppp \\
%\delta\psi &=& \frac{2}{\rho}\psi'.
%\eeqar

\section{Further Symmetries of the Fluid Model}
As mentioned above, since the fluid model descends from the
superstring, it should possess an enhanced symmetry beyond the Galileo
symmetry in (1+1)-dimensions.  In fact, the following conserved charges
effecting time rescaling and space-time mixing \cite{RJAPPRD} are
also verified:
\beqar
D &=& \int \rmd x \, (t{\cal H} - \rho\theta) \ , \\
G &=& \int \rmd x \, (x{\cal H} - \theta{\cal P}) \ ,
\eeqar

$G$ is sometimes referred to as the ``anti-boost'' because of its 
transformations on extended space-time \cite{Horvathy}.  The
Galileo generators supplemented by $D$ and $G$ together satisfy the
Lie algebra of the (2+1)-dimensional Poincar\'{e} group, with $N$, $P$, and $H$ corresponding to the three translations and with
$B$, $D$ and $G$ forming the (2+1)-dimensional Lorentz group $SO(2,1)$:
\begin{equation}
\begin{array}{ccc}
\big\{B,D\big\}= B,\hfill
&\big\{G,B\big\}=D,\hfill
&\big\{D,G\big\}=G \ ,
\end{array}
\end{equation}
with Casimir
\beqar
C = B \circ G + G \circ B + D \circ D \ .
\eeqar
Adjoining the supercharges results in the super-Poincar\'{e} algebra
of (2+1)-dimensions. The Lorentz charges do not belong to the infinite towers of constants of
motion mentioned earlier.  Rather, they act as raising and lowering
operators.  One verifies for the $Q_n$ and $I_n^{+}$:
\footnote{Note that the
$\{B,I_2^{+}\}$ bracket coincides with $\{B,2H\}$, which should equal
$-2P$ according to the Galileo algebra. But the above result,
viz. $-2I_1^{+}$, gives $-2(P-\sql\Omega)$.  This central addition arises from a term of
the form $$\int\rmd x \rmd y \sql \, x \frac{\partial}{\partial x} \delta(x-y),$$
whose value is ambiguous, depending on the order of integration.}
%\beqar
%\{I_n^{\pm},I_m^{+}\}=0,\qquad \{\bar{\eta}Q_n,\eta Q_m \}=\bar{\eta}\eta (I^{-}_{n+m}-2\sql nm A_{n+m-2}) \\
%\{A_n,A_m \}=?,\quad \{I_n^{\pm},Q_m \}=?,\quad \{I_n^{\pm},A_m \}=?, \quad \{A_n,Q_m \}=?  
%\eeqar
\begin{equation}
\begin{array}{ccc}
\big\{B,I_n^{+}\big\}=-n I_{n-1}^{+},\hfill
&\big\{D,I_n^{+}\big\}=(n-1) I_{n}^{+},\hfill
&\big\{G,I_n^{+}\big\}=(\frac{n}{2}-1) I_{n+1}^{+},\hfill
\\
\big\{B,Q_n\big\}=-n Q_{n-1},\hfill
&\big\{D,Q_n\big\}=(n-\frac12) Q_{n},\hfill
&\big\{G,Q_n\big\}=(\frac{n}{2}-\frac12) Q_{n+1},\hfill
\end{array}
\end{equation}
The brackets with the $I_n^{-}$ do not close, but the
$I_n^{-}$ can be modified by the addition of another tower of constant
quantities, namely those of (\ref{Aeq}):
\beq
\label{Itildeeq}
\tilde{I}_n^{-} = I_n^{-} - \sql n(n-1) \int \rmd x
R_{-}^{n-2}\psi\partial_x\psi \ .
\eeq
The modified constants obey the same algebra as $I_n^{+}$
\begin{equation}
\begin{array}{ccc}
\big\{B,\tilde{I}_n^{-}\big\}=-n \tilde{I}_{n-1}^{-},\hfill
&\big\{D,\tilde{I}_n^{-}\big\}=(n-1) \tilde{I}_{n}^{-},\hfill
&\big\{G,\tilde{I}_n^{-}\big\}=(\frac{n}{2}-1) \tilde{I}_{n+1}^{-}.
\end{array}
\end{equation}
Evidently $I_n^{+}$, $\tilde{I}_n^{-}$, and $Q_n$ provide irreducible, infinite dimensional representations for $SO(2,1)$, with the Casimir, in adjoint action, taking the form $l(l+1)$, and $l=1$ for $I_n^{+}$, $\tilde{I}_n^{-}$, and $l=1/2$ for $Q_n$.

Finally we inquire about the algebra of the towers of extended charges
$I_n^{+}$, $\tilde{I}_n^{-}$, and $Q_n$.  While some (bosonic) brackets vanish,
others provide new constants of motion like those in
(\ref{Ibeq})-(\ref{Keq}) and their generalizations with more
derivatives.  Thus it appears that one is dealing with an open (super) algebra.

%That this is
%an appropriate name can be seen even more clearly from the complete
%%algebra for the (1+1)-dimensional supersymmetric Chaplygin gas.  The
%notation of Riemann coordinates allows the full symmetry to manifest
%itself in the towers of conserved charges displayed in
%(\ref{bocharges}), (\ref{supercharges}), and (\ref{Acharges}).  Along with
%the boosts, dilations, and anti-boosts, we have:

%\beqar
%\{I_n^{\pm},I_m^{+}\}=0,\qquad \{\bar{\eta}Q_n,\eta Q_m \}=\bar{\eta}\eta (I^{-}_{n+m}-2\sql nm A_{n+m-2}) \\
%\{A_n,A_m \}=?,\quad \{I_n^{\pm},Q_m \}=?,\quad \{I_n^{\pm},A_m \}=?, \quad \{A_n,Q_m \}=?  
%\eeqar
%\begin{equation}
%\begin{array}{ccc}
%\big\{B,I_n^{\pm}\big\}=-n I_{n-1}^{\pm},\hfill
%&\big\{D,I_n^{\pm}\big\}=(n-1) I_{n}^{\pm},\hfill
%&\big\{G,I_n^{\pm}\big\}=(\frac{n}{2}-1) I_{n+1}^{\pm}+\frac{\sql}{2}n(1 \mp 1) A_{n-1},\hfill
%\\
%\big\{B,Q_n\big\}=-n Q_{n-1},\hfill
%&\big\{D,Q_n\big\}=(n-\frac12) Q_{n},\hfill
%&\big\{G,Q_n\big\}=\frac12(n-1) Q_{n+1},\hfill
%\\
%\big\{B,A_n\big\}=-n A_{n-1},\hfill
%&\big\{D,A_n\big\}=(n+1) A_{n},\hfill
%&\big\{G,A_n\big\}=(\frac{n}{2}+1) A_{n+1},\hfill
%\end{array}
%\end{equation}

%This is not just the (2+1)-dimensional super-Poincar\'{e} algebra, is
%it?

\section{Conclusions}
We have presented an integrable, supersymmetric fluid model with additional,
``dynamical'' symmetry tracing back to its origin in the superstring.
Besides the planar case in \cite{RJAPPRD}, this is the only other
dimensionality for a supersymmetric Chaplygin gas that can be obtained by going to the
light-cone gauge in a super $p$-brane.  

%Another parameterization is
%known to lead to the relativistic fluid Born-Infeld model from
%bosonic $p$-branes \cite{RJAP1}, and this is being studied with
%supersymmetry \cite{ongoing}.  

It remains an open question what other fluid interactions can be
obtained from the rich factory of branes. For example, string theory
$D$-branes have gauge fields living on them.  Such gauge fields would
presumably remain in passing to a fluid model and may thus provide a
model of magnetohydrodynamics from $D$-branes.  It might also be worthwhile to
explore whether the fluid models with Grassmann variables are suited
to describing the physical properties of fluids with spin degrees of
freedom.

\section*{Acknowledgments}
This work is supported in part by funds provided by the U.S.~Department of
Energy (DOE) under contract \#DE-FC02-94ER40818. One of the authors
(Y.B.) is also supported by a National Science Foundation (NSF) Graduate
Research Fellowship. 

% A useful Journal macro
\def\Journal#1#2#3#4{{#1} {\bf #2}, #3 (#4)}
\def\add#1#2#3{{\bf #1}, #2 (#3)}
\def\Book#1#2#3#4{{\em #1}  (#2, #3 #4)}
\def\Bookeds#1#2#3#4#5{{\em #1}, #2  (#3, #4 #5)}
% \Journal{}{}{}{}
% \Book{}{}{}{}

% Some useful journal names
\def\NPB{{ Nucl. Phys.}} % put <B> in next field
\def\PLA{{ Phys. Lett.}} % put <A> in next field
\def\PLB{{ Phys. Lett.}} % put <B> in next field
\def\PRL{{ Phys. Rev. Lett.}}
\def\PRD{{ Phys. Rev. D}}
\def\PR{{ Phys. Rev.}}
\def\ZPC{{ Z. Phys.} C}
\def\SJNP{{ Sov. J. Nucl. Phys.}}
\def\AnnP{{ Ann. Phys.}\ (NY)}
\def\JETPL{{ JETP Lett.}}
\def\LMP{{ Lett. Math. Phys.}}
\def\CMP{{ Comm. Math. Phys.}}
\def\PTP{{ Prog. Theor. Phys.}}
\def\PNAS{{ Proc. Nat. Acad. Sci.}}


\begin{thebibliography}{99}
\bibitem{comment}
The connection between the Nambu-Goto model in (3+1)-dimensional spacetime and the planar Chaplygin gas was noted by J. Goldstone and furher established by J. Hoppe (and collaborators). For a review see Ref.~\cite{RJAL}.
\end{thebibliography}
\end{document}